\documentclass[referee]{raa}            

\usepackage{graphicx,times}             

\begin{document}

   \title{The SAGE Photometric Sky Survey: Technical Description}

   \volnopage{Vol.0 (2018) No.0, 000--000}      
   \setcounter{page}{1}          

   \author{Zheng Jie
      \inst{1,2}
   \and Zhao Gang
      \inst{1,2}
   \and Wang Wei
      \inst{1,3}
   \and Fan Zhou
      \inst{1}
   \and Tan Ke-Feng
      \inst{1}
   \and Li Chun
      \inst{1}
   \and Zuo Fang
      \inst{1}
   }

    \institute{Key Laboratory of Optical Astronomy, National Astronomical Observatories,
	Chinese Academy of Sciences, Beijing 100101, China; {\it gzhao@nao.cas.cn}\\
    \and
    School of Astronomy and Space Science, University of Chinese Academy of Sciences, Beijing 100049, China\\
    \and
    Chinese Academy of Sciences South America Center for Astronomy,
	China-Chile Joint Center for Astronomy,
	National Astronomical Observatories, Chinese Academy of Sciences, Beijing 100101, China\\
   }

   \date{Received~~2018 Apr 29; accepted~~2018 May 25}

\abstract{  To investigate in more details of Stellar Abundance and Galactic Evolution (SAGE) and in a huge sample, we are performing a northern sky photometric survey named SAGES with the SAGE photometric system, which consists of 8 filters: Str\"omgren-$u$, SAGE-$v$, SDSS $g$, $r$, $i$, DDO-$51$, $H\alpha_{wide}$, and $H\alpha _{narrow}$, including three Sloan  broadband filters, three intermediate-band filters and two narrow-band filters, and one newly-designed narrow-band filter. SAGES covers $\sim$12,000 square degrees of the northern sky with $\delta > -5 ^{\circ}$, excluding the Galactic disk ($|b|<10^{\circ}$) and the sky area of 12 hr $<$ R.A. $<$ 18\,hr. The photometric detection limit depth at signal-to-noise ratio $5\sigma$ can be as deep as $V\sim$20\,mag. The SAGES will produce a depth-uniformed photometric catalogue for $\sim$500 million stars with atmospheric parameters including effective temperature $T_{\rm eff}$, surface gravity log\,g, and metallicity [Fe/H], as well as interstellar extinction to each individual target. In this work, we will briefly introduce the SAGE photometric system, the SAGE survey, and a preliminary test field of the open cluster NGC\,6791 and around.
\keywords{
	methods: observational
--- techniques: photometric
--- surveys
--- astrometry
--- catalogues }
}

   \authorrunning{ Zheng J., Zhao G., Wang W., Fan Z., Tan K.-F., Li C., \& Zuo F. }    
   \titlerunning{The SAGE Photometric Sky Survey: Technical Description }  

   \maketitle

\newcommand{\su}{$u_{\rm SC}$}
\newcommand{\sv}{$v_{\rm SAGE}$}
\graphicspath{{figure_pdf/}}

\section{Introduction} \label{sec:intro}

Astronomy is a basic science highly relying on the development of observations, and sky surveys on large  areas are the most important ladders for the progress of astronomy. In recent decades, SDSS (\cite{sdss}), 2MASS (\cite{2mass}) and other sky surveys have proved that mass data are very important, which lead to numerous new objects, new events, and new physics, and some even open a new area of modern astronomy. Besides photometric sky survey, spectrum surveys are also fruitful, for example LAMOST (\cite{lamost}) provides the most massive spectrum in the world, and plays an important role in finding metal-poor stars (\cite{lamost1}), high- or hyper-velocity stars (\cite{lamost3}), white dwarfs (\cite{lamost2}), emission line objects (\cite{lamost4}).

As far as we know, the Geneva-Copenhagen Survey (GCS, \cite{gcs}) is the only sky survey on the Str\"omgren-Crawford (SC) system (\cite{uvby63}, \cite{uvby64}, \cite{hbeta70}), which includes intermediate and narrow band filters, dedicated to stellar atmospheric parameters. The GCS survey is volume complete to a distance of 40\,pc down to $V\sim8.5$ mag. The HM catalogue (\cite{hm}) collected all measurements in these systems, which include only $\sim$66,000 stars in total. This situation is mainly due to the fact that it requires much more integration time in the narrow and medium-band, especially in the Str\"omgren-$u$ and $H\beta_{n}$-bands, than broadband filters, to achieve similar brightness limit. With the development of CCD photometry, a much better sensitivity in the blue of CCD detectors, and the emergence of tens of 4-10m telescopes, it is now a best opportunity to conduct a sky survey with a number of small infrequently-used smaller aperture telescopes with wide-field cameras in narrow and medium bandpasses which is dedicated to the determinations of atmospheric parameter of stars much fainter than GCS.

However, among all the previous or currently ongoing sky surveys, few of them are specifically focused on stellar atmospheric parameters. Under this situation, we are performing a deep photometric sky survey  using a combination of narrow, intermediate and broadband filters, with the main purpose to determine stellar parameters for hundreds of millions of stars with accuracies significantly better than those obtained with broadband filters and comparable to those determined spectroscopically. We will introduce the design of our sky survey and observations, and the data reduction pipeline, and our performance test field as a sample.

\section{The SAGE Photometric Sky Survey and Observation} \label{sec:sagepss}

\subsection{The SAGE Photometric System} \label{subsec:sagesystem}

The Johnson UBV photometric system (\cite{ubv}) is one of the earliest and most used among all standard photoelectric photometric systems. While the revolutionary project SDSS, by providing an unprecedented database of photometric observations of stars and galaxies, has essentially made its bandpasses, the SDSS photometric system, u'g'r'i'z' (\cite{sdssfilter}), the current standard for most (if not all) ongoing and future photometric surveys and most photometric imaging. However, it is unfortunate that SDSS filters were chosen basically to determine photometric redshifts for galaxies, rather than to isolate relevant stellar absorption features in particular bands, and therefore this system is not ideal for accurate determinations for the metallicities and gravity of stars.

Thanks to narrower bandpasses and specific definition in $u$ band, the Str\"omgren-Crawford (SC) system  provides a reliable method to determine stellar parameters for stars with a wide range of spectral types (e.g., early or late type stars, metal-poor stars). In particular, this photometric system, composed of the Stro\"mgren $ubvy$ plus the H$\beta$ wide and narrow bandpasses, can accurately identify stars at various evolutionary stages (\cite{uvby63, sc2010}). Several indices, including $m1=(v-b){-}(b-y)$, $c1=(u-v){-}(v-b)$, $\beta=\beta_{n}-\beta_{w}$ could be used to measure metallicity and surface gravity, and the strength of H$\beta$ line without being affected by extinction.

\begin{figure}[htbp]
\begin{center}
\includegraphics[width=\textwidth]{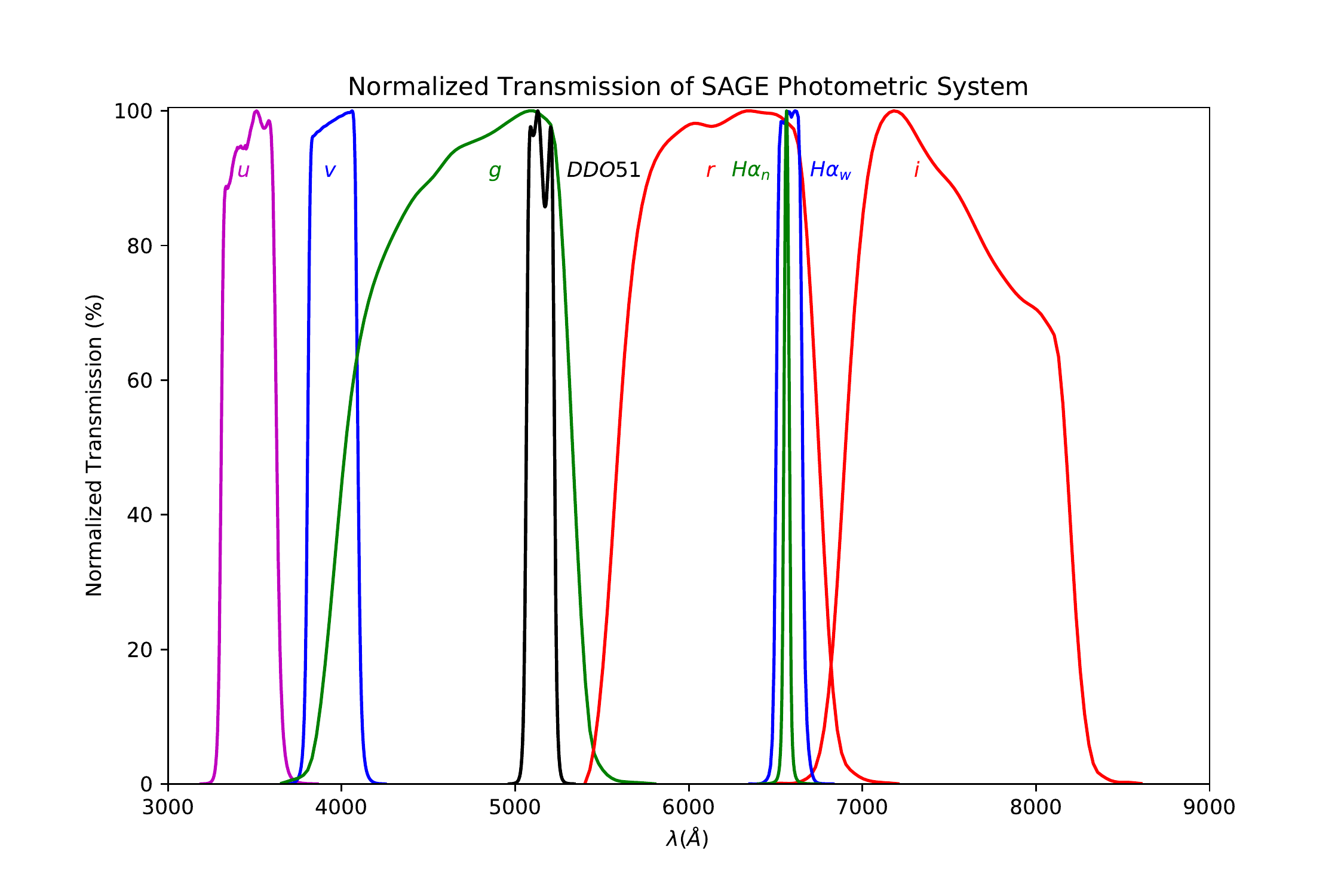}
\caption[The Normalized Transmission Curves of the SAGE Filters]{The Normalized Transmission Curves of the SAGE Filters} \label{fig:SageSystem}
\end{center}
\end{figure}

In order to obtain accurate stellar atmosphere parameters with high efficiency, we constructed a new photometric system: the SAGE system, by combining multiple photometric systems, and adding several new filters. This system consists of 8 filters: Str\"omgren-$u$, SAGE-$v$, SDSS $g$, $r$, $i$, DDO-$51$, $H\alpha_{wide}$, and $H\alpha _{narrow}$ (hereafter \su, \sv, $g$, $r$, $i$, DDO$51$, $H\alpha_{w}$ and $H\alpha_{n}$,  respectively). The effective wavelengths and bandwidths of them are shown in Table~\ref{tab:SageSystem} and their normalized transmission curves are plotted in Fig.~\ref{fig:SageSystem}. All magnitudes of the SAGE system are on the AB system.

The SAGE system can effectively determine stellar atmospheric parameters, including effective temperature $T_{\rm eff}$, surface gravity log\,g, and metallicity [Fe/H], as well as extinction. These parameters can be determined by $(g-i)$, $(u_{\rm SC} {-} v_{\rm SAGE})$, $(v_{\rm SAGE}-g)$, and $(H\alpha_{n}-H\alpha_{w})$ in the system. The response curves of the \su{} and \sv{} filters are located on both sides of the Balmer jump and can be used to measure its intensities. Moreover, there is almost no overlapping area between the two filters of the SAGE system, and more accurate log\,g measurement results can be obtained. We prefer our \sv{} than Str\"omgren-$v$ as it is bluer by $\sim$150\AA, aiming to include Ca~{\sc ii} H \& K doublets for better correlation with stellar metallicities. The SDSS $g$ and $r$ filters should have similar usages as Str\"omgren $b$ and $y$, while they will consume significantly less amount of observation time with the latter two. We change the H$\beta$ filters to H$\alpha$ ones as the CCD efficiency is similar at both wavelengths, while the latter ones have stronger imprints on stellar or nebular spectra, and therefore should have higher sensitivity. In G- and K-type stars, H$\beta$ lines are weaker than H$\alpha$ ones, so measurement on these lines will be more efficient. The DDO51 filter is employed to further constrain stellar gravity. For more details, please refer to \cite{fan18}.

\begin{table}[htbp]
\caption[The SAGE Photometic System]{The SAGE Photometic System} \label{tab:SageSystem}
\begin{center}
\begin{tabular}{lcccccccc}
\hline\noalign{\smallskip}
 Band & \su & \sv & $g$ & $r$ & $i$ & DDO51 & $H\alpha_n$ & $H\alpha_w$ \\
\hline\noalign{\smallskip}
Effective wavelength (\AA) & 3520 & 3950 & 4639 & 6122 & 7439 & 5132 & 6563 & 6563 \\
Bandwidth            (\AA) &  314 &  290 &1280 &1150 &1230 &  162 &   29 &  136 \\
\noalign{\smallskip}\hline
\end{tabular}
\end{center}
\end{table}

\subsection{The SAGE Photometric Sky Survey}\label{subsec:sagesurvey}

The SAGE Photometric Sky Survey (SAGES) is a northern sky survey with the SAGE photometric system on about 12,000 square degrees. This survey was firstly proposed in 2014 (\cite{wang14}) to be conducted with the full set of the Str\"omgren-Crawford (SC) system. After further investigations, we believe our SAGE system is superior to the SC system, and the SAGE system is therefore adopted in our survey instead of the SC system. Since 2015, the SAGES has been started, and the plan is to finish observations, flux calibrations and astrometric calibrations in 4 to 5 years. We aim to achieve a  survey depth  at signal-to-noise ratio (SNR) 5:1 to 20 mag in the Johnson $V$ band on Vega system, which corresponds to  depths of $\sim$21.5 in \su, $\sim$21.0 in \sv, and $\sim$19.5 in $g$, $r$ and $i$ . Note that these are magnitudes on the AB system, and all flux and errors are measured from point sources extracted by Source Extractor (\cite{sex}) with aperture photometry with an aperture diameter of 2.5 times the mean fwhm of objects on each image. The SAGES will produce a depth-uniformed photometric catalogue for $\sim$500 million stars with atmospheric parameters including effective temperature $T_{\rm eff}$, surface gravity log\,g, and metallicity [Fe/H], as well as extinction to each individual target. Value-added information like stellar radius may be provided as well when combined with high precision parallax measurements by the Gaia Mission (\cite{gaiadr2}).

To finish a whole northern sky survey with non-broad band filters takes a lot of telescope time even with wide-field cameras. Therefore, we decide to use three telescopes almost simultaneously to conduct the survey in different bandpasses, as described below. We note that with careful design of standard star observations, the absolute flux calibration is expected to be as good as 0.02\,mag, and the uniformity to be better than 0.02\,mag in our first data release, and shall be improved in later data releases.

\subsection{Telescopes and Instruments} \label{subsec:tel}

Given that the \su-band has a wavelength coverage very close to the atmospheric cutoff, plus the fact in this band typical stars are intrinsically very faint, therefore integration time in \su{} to achieve our proposed $5\sigma$-limit of 21.5\,mag. Considering to keep the observations in three different telescopes at the same pace, and the fact the Bok 90-inch telescope\footnote{http://james.as.arizona.edu/$\sim$psmith/90inch/90inch.html} has twice larger aperture than the other two, we decide to conduct the observations in \su{} and \sv. The Bok telescope is an equatorial mounting telescope located at Kitt Peak, which is at $30^{\circ}57'46''.5$N by $111^{\circ}36'01''.6$W, and 2071 meters above sea level. The Bok Telescope belongs to Steward Observatory, the University of Arizona.

\begin{figure}[htbp]
\begin{center}
\includegraphics[width=0.75\textwidth]{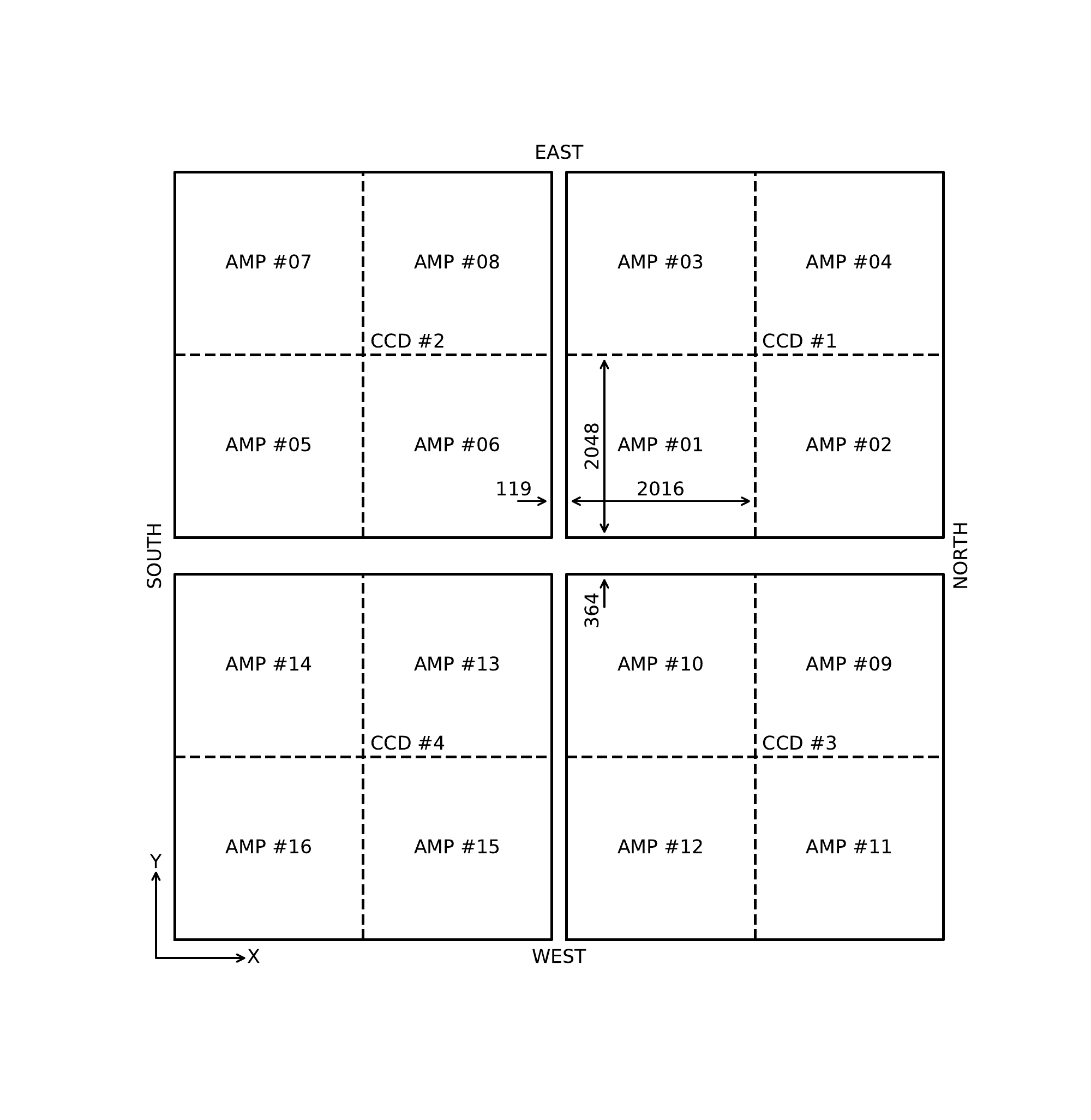}
\caption[The layout of 90Prime detector array]{The layout of 90Prime detector array. (Not to Scale, Pixel)} \label{fig:boklayout}
\end{center}
\end{figure}

The instrument named 90Prime consists of four 4K $\times$ 4K back-illumination CCDs and is installed at the prime focus of the telescope. The field of view is around $1^{\circ}.08 \times 1^{\circ}.03$ with gaps of $\sim166''$ at R.A. and $\sim54''$ at Decl between CCDs. Fig.~\ref{fig:boklayout} shows the layout of 90Prime. We choose the slow mode to read the whole frame of 90Prime, which takes about 37 seconds with a noise level of 6-10 electrons per pixel. We perform \su- and \sv-band observations using the Bok telescope.

For the SDSS $g$, $r$ and $i$-bands, we decide to use Nanshan One-meter Wide-field Telescope (hereafter, NOWT) to carry out observations. NOWT is an altazimuth-mounted telescope located at Nanshan Station, which is at $43^{\circ}16'45''.0$N by $87^{\circ}10'38''.3$E, and 2081 meters above sea level. NOWT belongs to Xinjiang Astronomical Observatory (XAO), Chinese Academic of Sciences (CAS). A wide field camera, consisting of a 4K$\times$4K blue-enhanced back-illumination CCD, is installed at the prime focus of NOWT, with a field of view of about $1^{\circ}.5 \times 1^{\circ}.5$. The typical readout speed is about 40 seconds via 4 parallel amplifiers with a readout noise of 8-10 electrons (\cite{nowt}). As SDSS survey has covered $\sim$9,000 square degrees in the northern sky, we only need to conduct the observations in $gri$-bands for the rest $\sim$4,000 square degrees including some overlapping regions to secure flux calibrations.

The Zeiss-1000 Telescope at the Maidanak Astronomical Observatory, Ulugh Beg Astronomical Institute, Uzbek Academy of Sciences (MAO, \cite{mao}) is being upgraded, it is expected to be available in late 2018. Its current field of view is $32.9' \times 32.9'$. We will perform $H\alpha_{\rm w}$ and $H\alpha_{\rm n}$ observations there.

\subsection{Observations and progress} \label{subsec:obs}

\begin{figure}[htbp]
\centering
\includegraphics[width=0.75\textwidth]{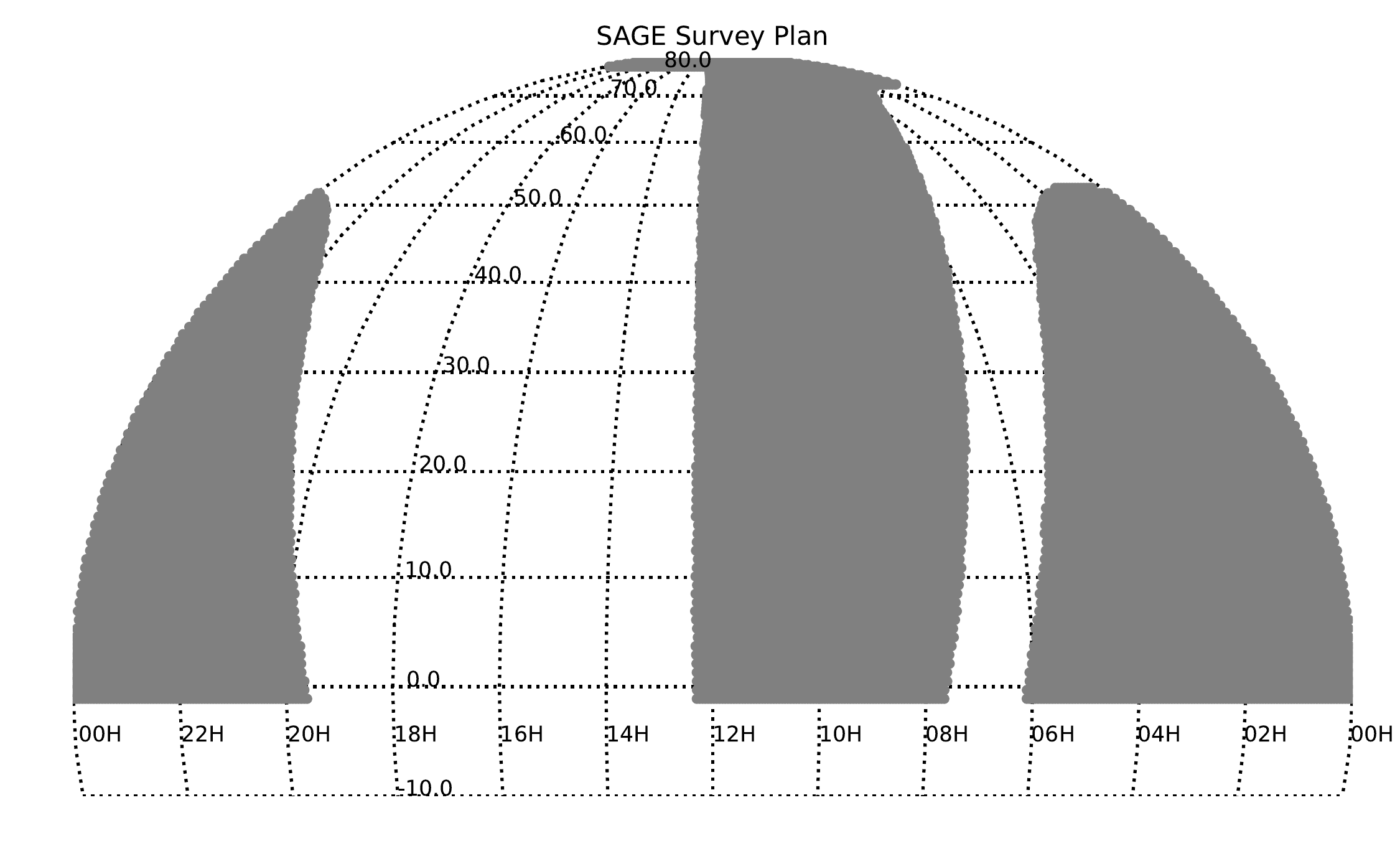}
\caption[Coverage Plan of the SAGE photometric Sky Survey]{Coverage Plan of the SAGE photometric Sky Survey}\label{fig:sageplan}
\end{figure}

The SAGES is proposed to cover the northern sky with $\delta > -5 ^{\circ}$, excluding the bright, high extinction Galactic disk ($|b|<10^{\circ}$). Meanwhile, because of the time allocation of Bok, we exclude the sky area of 12 hr $<$ R.A. $<$ 18\,hr for the current observation, and may accomplish it in future projects. Fig.~\ref{fig:sageplan} demonstrates the proposed coverage of the SAGES.

For all the observations at the three telescopes, we have the same strategy with the only differences are filters and exposure times. In clear nights, the exposure time for each band is listed in Table~\ref{tab:exptime}. The exposure times will increase as the  airmass increases. They will also increase with the cloud. For each night we make a plan based on the previous footprint, and the current date, so that most fields are observed at lowest possible airmass, and the time for telescopes to slew is the shortest. Normally, a field will be observed only once for each band.

\begin{table}[htbp]
\caption[Exposure time of each band]{Exposure time of each band in clear nights} \label{tab:exptime}
\begin{center}
\begin{tabular}{cccccc}
\hline\noalign{\smallskip}
Band          & \su & \sv & $g$ & $r$ & $i$  \\
\hline\noalign{\smallskip}
Exposure time (second) & 60 & 20 & 30 & 40 & 40 \\
\noalign{\smallskip}\hline
\end{tabular}
\end{center}
\end{table}

\begin{figure}[htbp]
\centering
\includegraphics[width=0.75\textwidth]{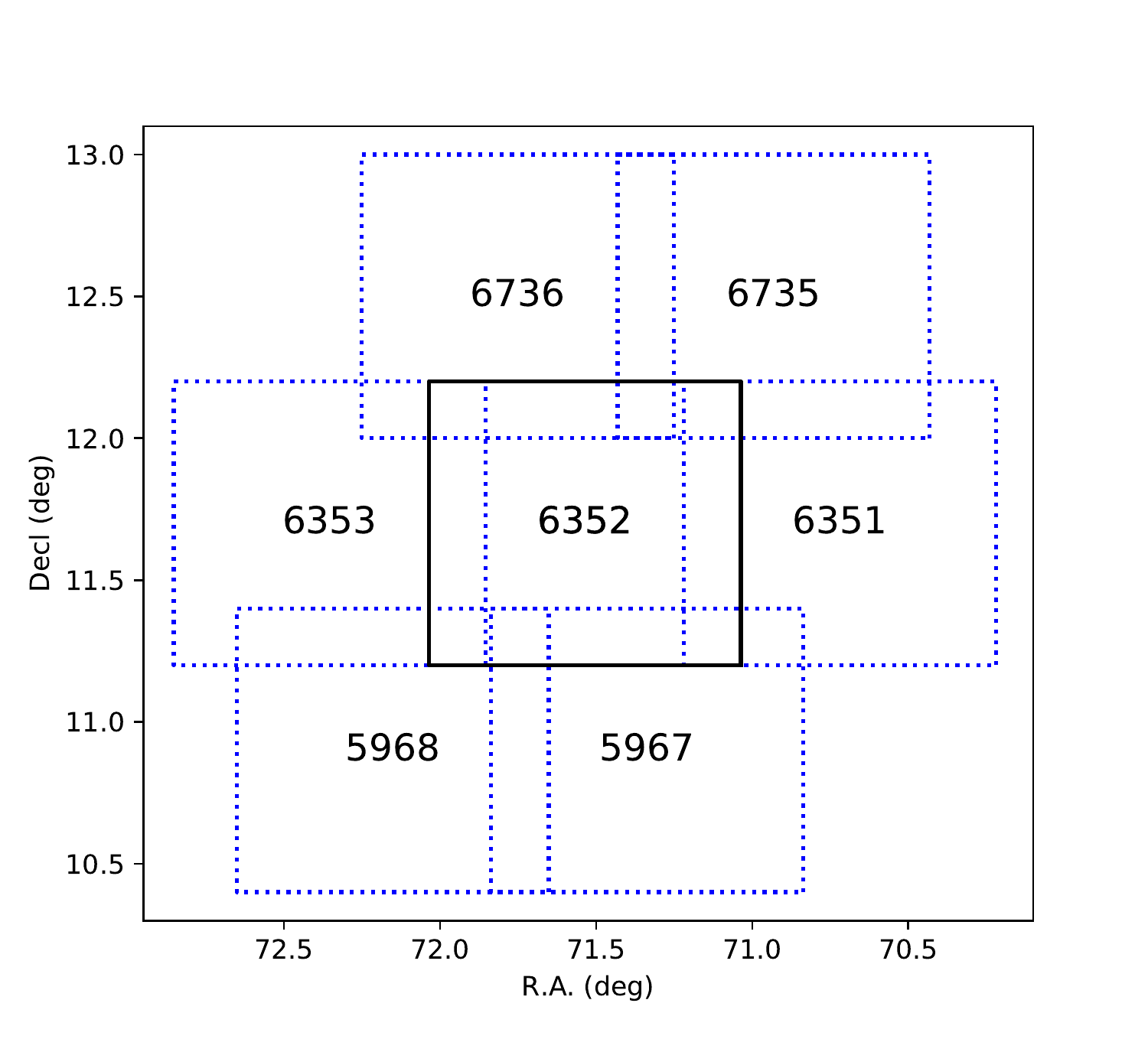}
\caption[Overlap between Survey Fields]{Overlap between Survey Fields}
\label{fig:overlap}
\end{figure}

To achieve reliable and consistent flux and astrometric calibrations, we leave sufficient overlapping regions between adjacent fields. The observation fields at the same Declination form a horizontal stripe, and in the stripe $\sim$20\% overlapping is reserved between adjacent fields in the same stripe. Meanwhile $\sim$20\% overlapping is set between adjacent stripes. As a result, each field will have one neighbour in the east and one in the west with $\sim$20\% overlapping region in each direction, and it has two northern and two southern neighbours with overlapping fractions of 0-20\% for each field-to-field pair, but in total 20\% for each direction. Fig.~\ref{fig:overlap} shows the overlapping between Bok Field 6352 and its neighbours as an example, and the distributions are very similar with different field scales due to different size fields-of-view for the other two telescopes. We note that the Fig.~\ref{fig:overlap} does not show the internal gaps between mosaic CCDs of Bok.

We estimate that $\sim$36\% of the survey area will be visited only once, and $32-48$\% of survey area will be visited twice, and $\sim$24\% to be visited three times, and the rest small fraction $<$16\% to be visited four times. The time span between visit could range from minute to days and to years. With multiple visits of in total 36\% of our survey area, on one hand, reliable flux and astrometric calibrations could in principle achieved, and on the other hand, a time-domain alert should be able to be spotted out for future follow-up studies.

The SAGES launches its observations in the autumn of 2015. By the end of Jan 2018, observations of $g$, $r$ and $i$ at NOWT are completed, meanwhile, observation of \su{} and \sv{} on Bok are about $2/3$ completed, and the rest $1/3$ is expected to be done in 2019. Observations to be done at MAO is scheduled in late 2018. The current progress of the Survey as of January 2018 are listed in Table~\ref{tab:progress}, and the up-to-date progress can be found at the official website of the SAGES\footnote{http://sage.bao.ac.cn/surveyobs/obsfootprint.php}.

\begin{table}[htbp]
\caption[The observation progress of the SAGES]{The observation progress of the SAGES as of January 2018.} \label{tab:progress}
\begin{center}
\begin{tabular}{cccccc}
\hline\noalign{\smallskip}
Band          & \su & \sv & $g$ & $r$ & $i$  \\
\hline\noalign{\smallskip}
Field number           & 12364 & 11376 &  4254 &  4254 &  4254 \\
Area covered (deg$^2$) &  7913 &  7280 &  4254 &  4254 &  4254 \\
Percent       (\%)     &  69.7 &  64.2 & 100.0 & 100.0 & 100.0 \\
\noalign{\smallskip}\hline
\end{tabular}
\end{center}
\end{table}


\section{The Data Reduction Pipeline} \label{sec:sdp}

\begin{figure}[htbp]
\begin{center}
\includegraphics[width=\textwidth]{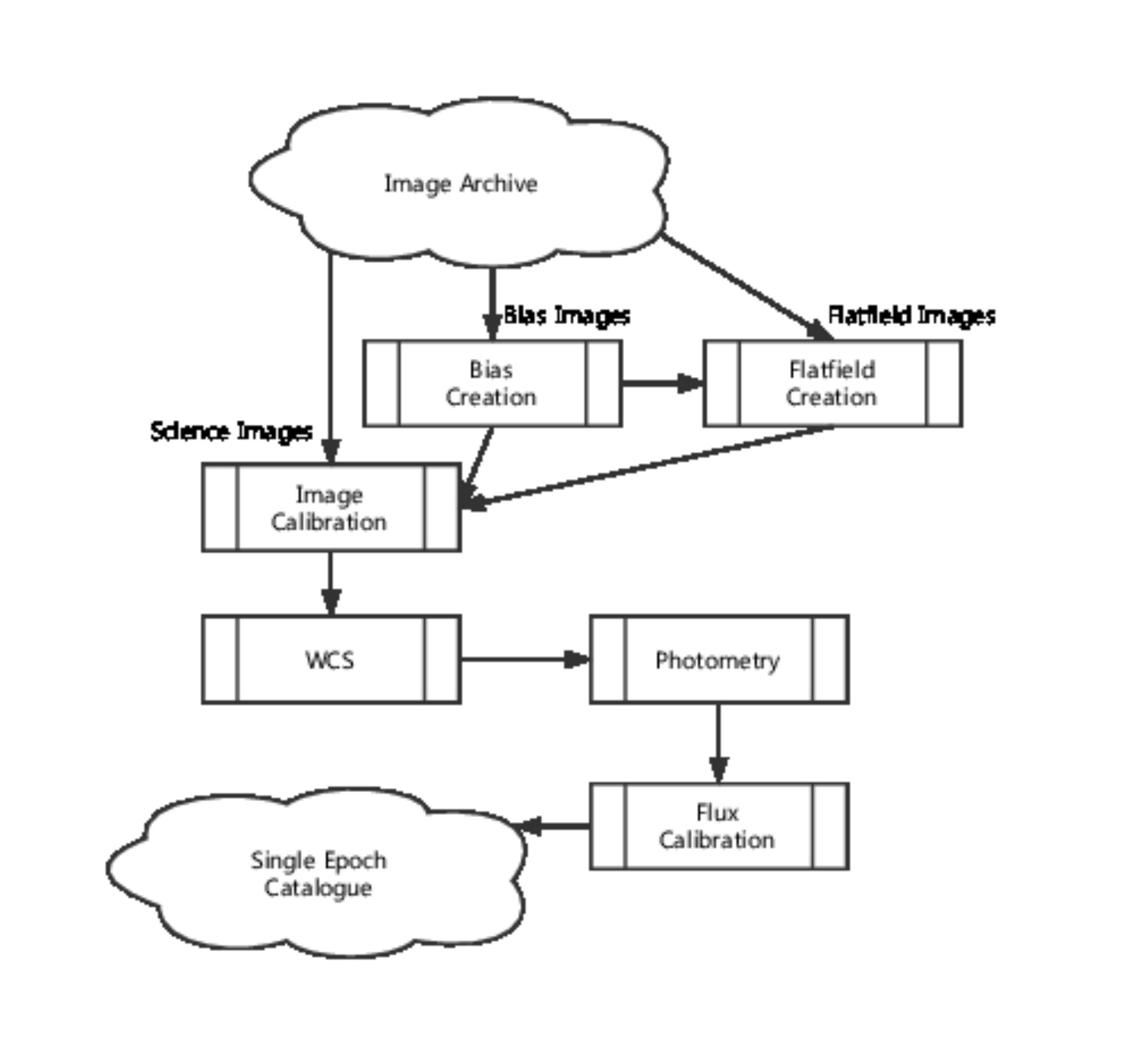}
\caption[The Flowchart of the Science Data Pipeline]{The Flowchart of the Science Data Pipeline}\label{fig:pipeline}
\end{center}
\end{figure}

We have developed a data reduction pipeline, to semi-automatically perform standard reduction procedures including bias subtraction, flat-fielding, distortion correction, astrometry, and aperture photometry.  We will introduce our pipeline briefly in this section, as shown in Fig.~\ref{fig:pipeline}.

\subsection{Correction of Images} \label{subsec:imagecorr}

The ultimate purpose of image reduction is to convert detector digital counts given in raw CCD images to electron count rates (e$^{-}$/s), through a series of procedures including overscan and bias correction, flat-fielding, and crosstalk removing.

\begin{figure}[htbp]
\begin{center}
\includegraphics[width=\textwidth]{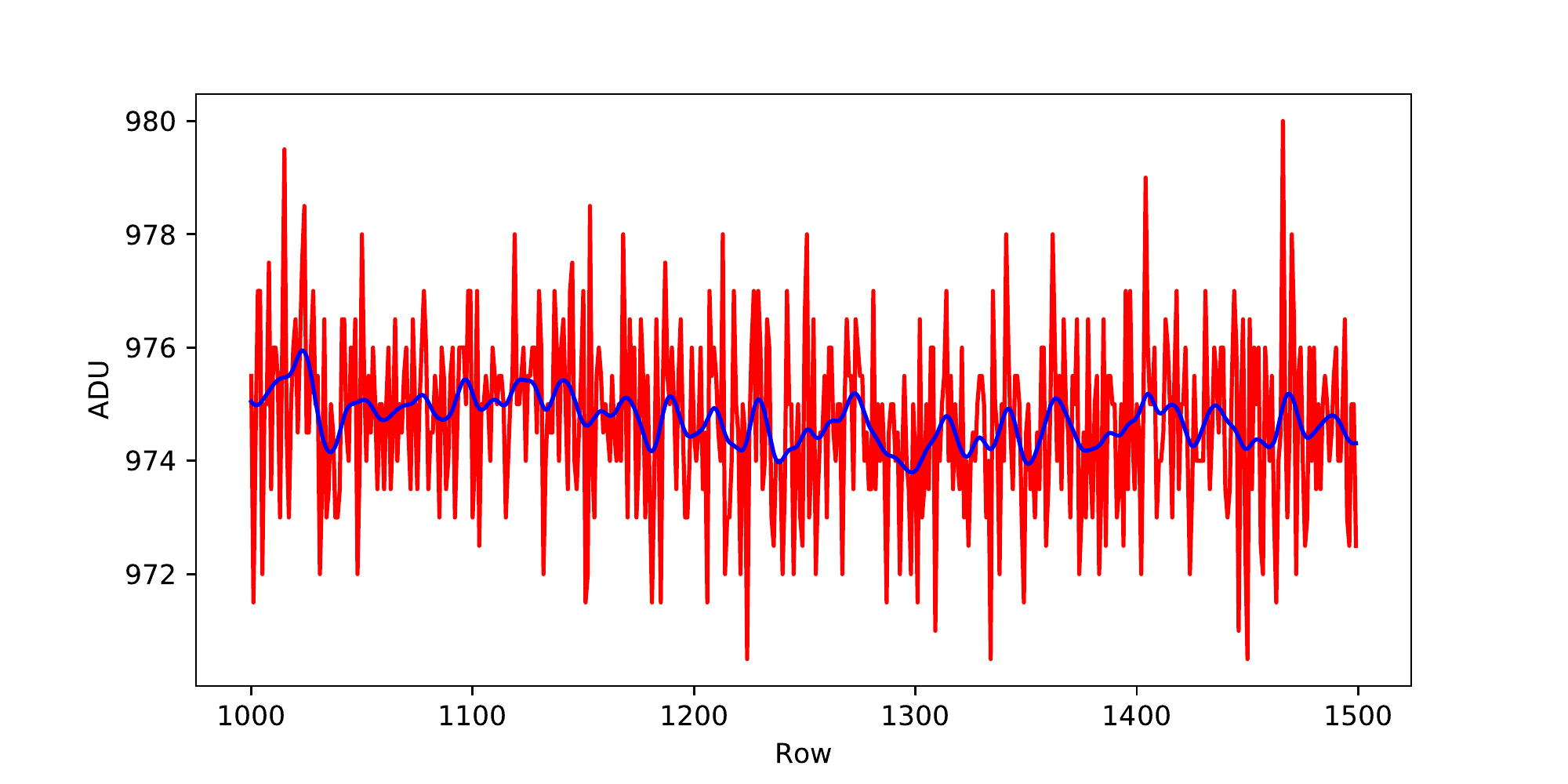}
\caption[ADU of overscan region and a smoothed curve]{ADU of overscan region (red) and a smoothed curve (blue)} \label{fig:overscan}
\end{center}
\end{figure}

The pipeline computes the median value of each row of the overscan area, then subtracts it from the corresponding rows of the image. Following the technique performed in \cite{bassdr1} for the BASS data, which was obtained also with 90Prime, a Gaussian smoothing is applied to the counts variations in the row direction, which is believed to better represent bias level along the row direction. A section of the overscan data and its smoothed result are displayed in Fig.~\ref{fig:overscan}. For all images, including bias, flat-field, and science images, the overscan correction is the first step in data reduction. We use the same method to correct overscan at both Bok and NOWT after tests.

Before and after each night of observations, a set of 10 bias exposures are taken. A master bias frame can be easily constructed, with each pixel's value to be the median value of the 20 exposures at the same pixels. With this master bias image, all science exposures and flat-field exposures taken on the same night can be bias-subtracted accordingly.

The next step is to perform flat-fielding for science exposures with flat-field exposures taken at the sky during twilight, or those taken on a curtain illuminated uniformly by a scanning lamp during the daytime, or super night-sky flat by merging all the science images taken over the night. As usual, the priority is to use high signal-to-noise ratio (SNR) twilight flats to correct large-scale illumination trend and employ the dome flats to correct for pixel-to-pixel variations when applicable, and if not, a super night-sky flat will be used instead. We estimate that our master flat images will be accurate to 1 percent for most nights.

The 90Prime consists of 4 CCDs, each has 4 amplifiers for parallel readout. This function of multi-channel CCD read-out can significantly reduce the time spent on reading the detector. The disadvantage of this option is the so-called amplifier cross-talk, which causes contamination across the output amplifiers, typically at the level of 1:10,000. It can a serious problem for high precision photometry, and when bright stars saturate in CCD detectors.

\begin{figure}[htbp]
\begin{center}
\includegraphics[width=\textwidth]{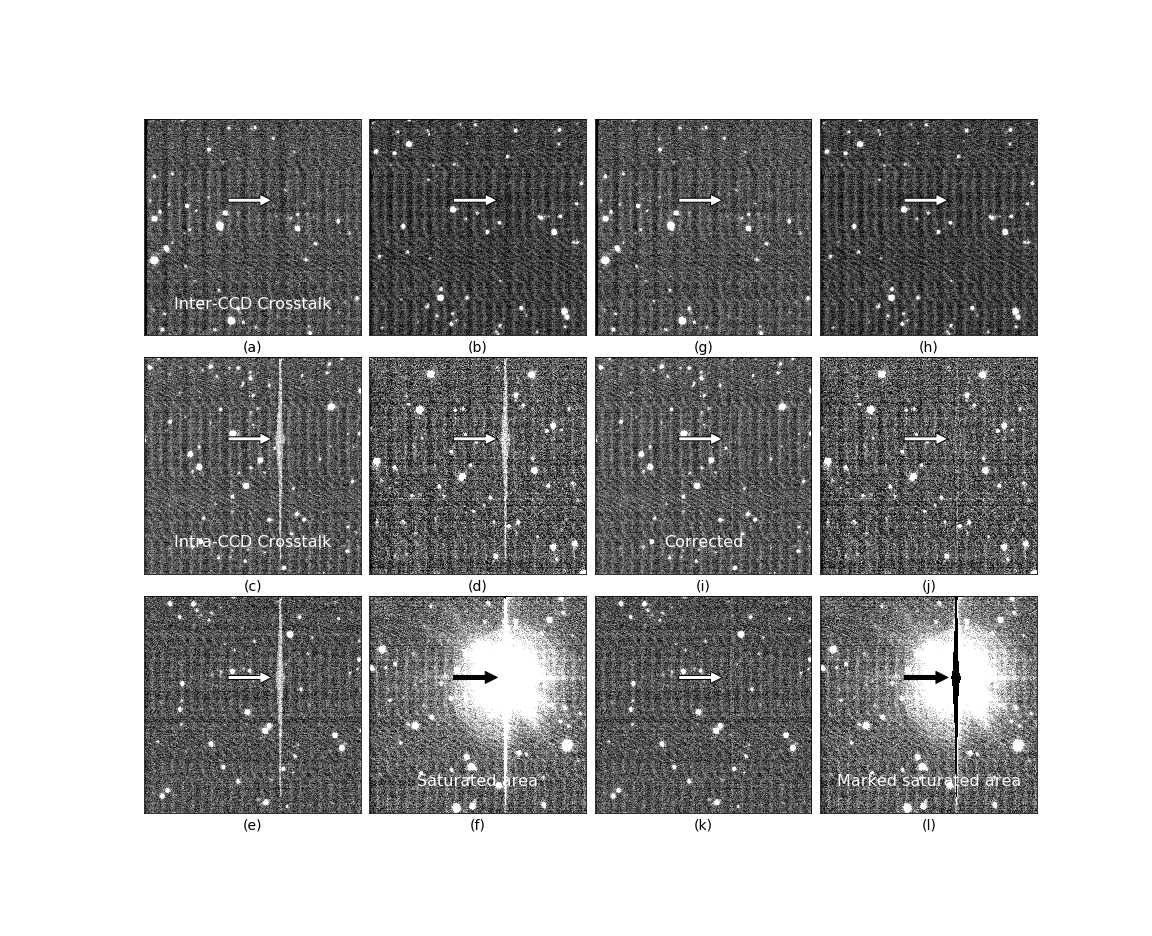}
\caption[Image sections before and after crosstalk correction]{A typical crosstalk and the image after correction in inverted gray scale. Panel a-f are original images. Panel f is the source, panel c-e are polluted by intra-CCD crosstalk, and panel a-b are polluted by inter-CCD crosstalk. Panel g-k are corrected images of a-e, respectively. In panel j, the saturated area is marked. In all panels, the polluted and corrected areas are pointed out. Stripes caused by interference from the camera controllers as large-scale patterns can be seen in images. We are working on this and they will be removed in the future.}
\label{fig:crosstalk}
\end{center}
\end{figure}

We following the method suggested by \cite{crosstalk} to remove crosstalks.  We compute the crosstalk coefficient from bright sources detected and their crosstalk signal on other amplifiers of the image. We use a large number of images to estimate the crosstalk coefficients between amplifiers. The pipeline computes the crosstalk pollution from bright areas with the coefficients, then apply it to correct the image. The overall ratio of crosstalk is about 5:10,000 for 90prime, while inter-CCD crosstalk ratios are greater and intra-CCD ratios are lower. Fig.~\ref{fig:crosstalk} shows a typical crosstalk and the image after correction.  Panel a-f are original images. Panel f is the source, panel c-e are polluted by intra-CCD crosstalk, and panel a-b are polluted by inter-CCD crosstalk. Panel g-k are corrected images of a-e, respectively. In panel j, the saturated area is marked. In all panels, the polluted and corrected areas are pointed out.

In Fig.~\ref{fig:crosstalk}, stripes caused by interference from the camera controllers can be seen. They are large-scale patterns but not uniform and do not full-fill the whole image. We are working on this issue and they will be removed in the future version of the pipeline.

\subsection{Astrometric Calibration} \label{subsec:wcs}

The astrometric calibration is realized in two steps. Firstly, the pipeline will work out a linear solution assuming the image has no distortion, based on the information stored in image headers, including the telescope pointing coordinates, the rotation angle, and the pixel scale.

For further corrections, the pipeline employees the package named Software for Calibrating AstroMetry and Photometry (SCAMP, \cite{scamp}) to derive astrometric solutions. SCAMP cooperates with SExtractor (\cite{sex}) by matching SExtractor's output catalogue with an online or local reference catalogue. SCAMP is mature and robust software that is widely used in astrometric calibration. The pipeline calls twice SCAMP. For the first one, a loose criterion for cross-matching detected sources with reference sources is used to retain enough stars for calibration. For the second run of SCAMP, a strict criterion is used to obtain precise solution based on previous results.  Table~\ref{tbs:campconf} shows the major configuration parameters for SCAMP in the two calls. The configurations are same for both Bok and NOWT, especially we find DISTORT\_DEGREES = 3 is the most suitable configuration after a series of tests.

\begin{table}[htp]
\caption[The major configuration parameters for SCAMP]{The major configuration parameters for SCAMP.}\label{tbs:campconf}
\begin{center}
\begin{tabular}{llll}
\hline\noalign{\smallskip}
Keyword           & Round 1  & Round 2 & Note \\
\hline\noalign{\smallskip}
MATCH             &  Y       &   Y     &   Module match or not \\
MATCH\_NMAX       &  0       &   0     &   Up bound of cross match \\
PIXSCALE\_MAXERR  &  2       &   1.5   &   Max error of pixel scale \\
POSANGLE\_MAXERR  &  5.0     &   2.0   &   Max error of position angle (degrees) \\
POSITION\_MAXERR  &  10.0    &   1.0   &   max error of position (arcminutes) \\
MATCH\_RESOL      &  0       &   0     &   Matching resolving \\
MATCH\_FLIPPED    &  N       &   N     &   Allow axis flipping in match or not \\
CROSSID\_RADIUS   &  25.0    &   2.0   &   Cross identification radius (arcseconds) \\
SOLVE\_ASTROM     &  Y       &   Y     &   Solve astrometric solution or not \\
PROJECTION\_TYPE  &  SAME    &   SAME  &   Projection type \\
DISTORT\_DEGREES  &  3       &   3     &   Degrees of distortion polynomial \\
\noalign{\smallskip}\hline
\end{tabular}
\end{center}
\end{table}

The Position and Proper Motion Extended (PPMX, \cite{ppmx}) catalogue is adopted in our pipeline as the astrometric reference. PPMX contains about 18 million stars which are evenly distributed across the whole sky with accuracies of $\sim$0.02\arcsec{} in both R.A. and Decl. directions. Over 85\% PPMX stars have $V$ magnitudes between 10.0 and 15.0 mag, which matches very well with the dynamic ranges of our SAGES survey.  As the Gaia DR2 (\cite{gaiadr2}) has just been released, we will use it as the astrometric reference in the future.

SDSS and Pan-STARRS DR1 (PS1, \cite{panstarrs}) are both accurate in astrometric and flux calibrations. But SDSS covers only part of all sky, so we cannot use it as the reference catalogue to ensure the uniformity of the correction. PS1 has an uncertainty lower than 0.005\arcsec{} in both R.A. and Decl, but it does not provide proper motions, so we do not use it as the astrometric reference, but we use it as the flux reference in Section \ref{subsec:flux}.

Besides calling SCAMP, another astrometric calibration method is developed to provide an astrometric solution. Both methods are used in the pipeline so that we can compare their results to make sure we have the correct solution.

Generally, the purpose of the astrometric calibration is to fit a transforming formula from image coordinates $(x, y)$ to celestial sphere coordinates $(\alpha, \delta)$, and then apply the formula to obtain the coordinates of all detected objects. The transforming formula is usually linear functions if there are not significant image distortions. However, both the 90Prime and the NOWT camera are at prime focus with a fast focal ratio, and the field-of-views are both larger than 1 deg$^2$, the image distortions cannot be ignored and should be corrected.

The first step is to convert the original pixel coordinates $(x, y)$ with origin at the left-bottom corner, to intermediate pixel coordinates $(u, v)$ with the origin at the center of the image. Then $(u, v)$ is converted to intermediate world coordinates $(\xi , \eta)$ using parameters CDi\_j in the image header. Finally, $(\xi, \eta)$ is projected to the world coordinates $(\alpha, \delta)$, depending on the projection type, and our pipeline adopts the type `TAN'. To resolve non-linear correlation between $(\xi , \eta)$ and $(\alpha, \delta)$, we use SIP (Simple Imaging Polynomial, \cite{sip}) convention to represent image distortion. SIP adds high-order correcting polynomials $f$ and $g$ to $u$ and $v$ to express the distortion, as shown in formula \ref{equ:sip}.

\begin{equation}
\left( \begin{array}{c} \xi \\ \eta \\ \end{array} \right)
=
CD
\times
\left( \begin{array}{c} u + f(u,v) \\ v + g(u,v) \\ \end{array} \right)
\label{equ:sip}
\end{equation}

To determine polynomials $f$ and $g$, $A_{pq}$ and $B_{pq}$ are used as the coefficients of $u^pv^q$ as shown in Formula~\ref{equ:sipfg}, in which ${\rm N_A}$ and ${\rm N_B}$ are the highest order to correct $u$ and $v$, respectively. After comprehensive tests, we find ${\rm N_A} = {\rm N_B}= 3$ is appropriate for the SAGES.

\begin{equation}
\begin{array}{l}
f(u,v) = \sum_{p,q} {A_{pq} \cdot u^p v^q}, \quad 2 \le p+q \le {N_A} \\
g(u,v) = \sum_{p,q} {B_{pq} \cdot u^p v^q}, \quad 2 \le p+q \le {N_B}
\end{array}
\label{equ:sipfg}
\end{equation}

\begin{figure}[htbp]
\begin{center}
\includegraphics[width=\textwidth]{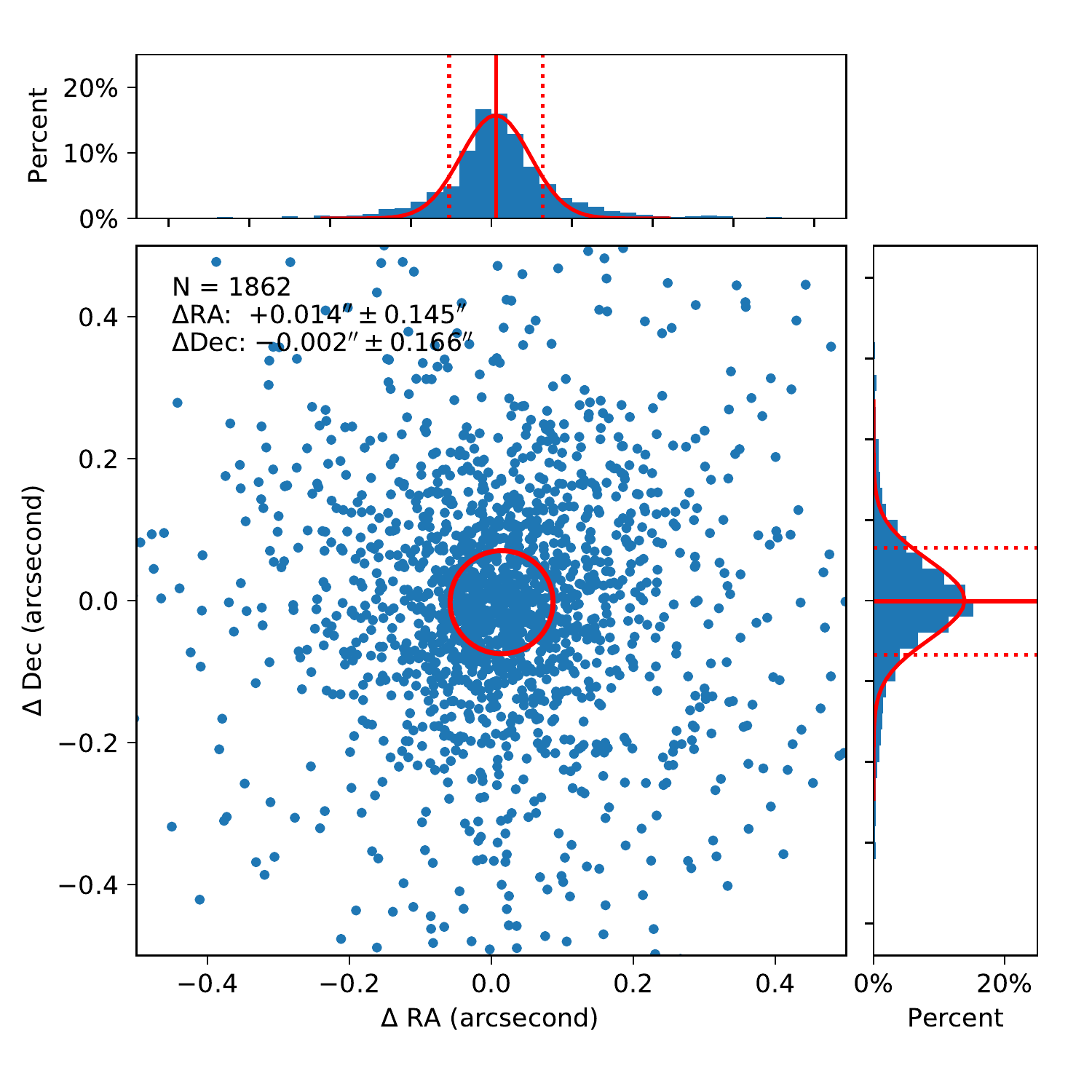}
\caption[A typical internal error of the astrometric results]{Typical internal error of the astrometric results yielded by the SAGES pipeline, as compared to PPMX. The number of stars used for comparisons, the offsets, and standard deviations are marked in the lower-left panel. In top and right panel, the histograms of the differences in R.A. and Decl are plotted, respectively, with best-fit Gaussian-profile overplotted in red.
} \label{fig:intwcserror}
\end{center}
\end{figure}

To evaluate internal astrometric errors yielded by the SAGES pipeline, comparisons have been made for all the fields with multiple visits, either in the same band, or in different bands, and calculate the differences in coordinates between different visits. Fig.~\ref{fig:intwcserror} shows a typical internal astrometric error, in which we find that $\Delta R.A. = 0''.014\pm0.''145$ and $\Delta Decl = -0''.002\pm0.''166$.

\begin{figure}[htbp]
\begin{center}
\includegraphics[width=\textwidth]{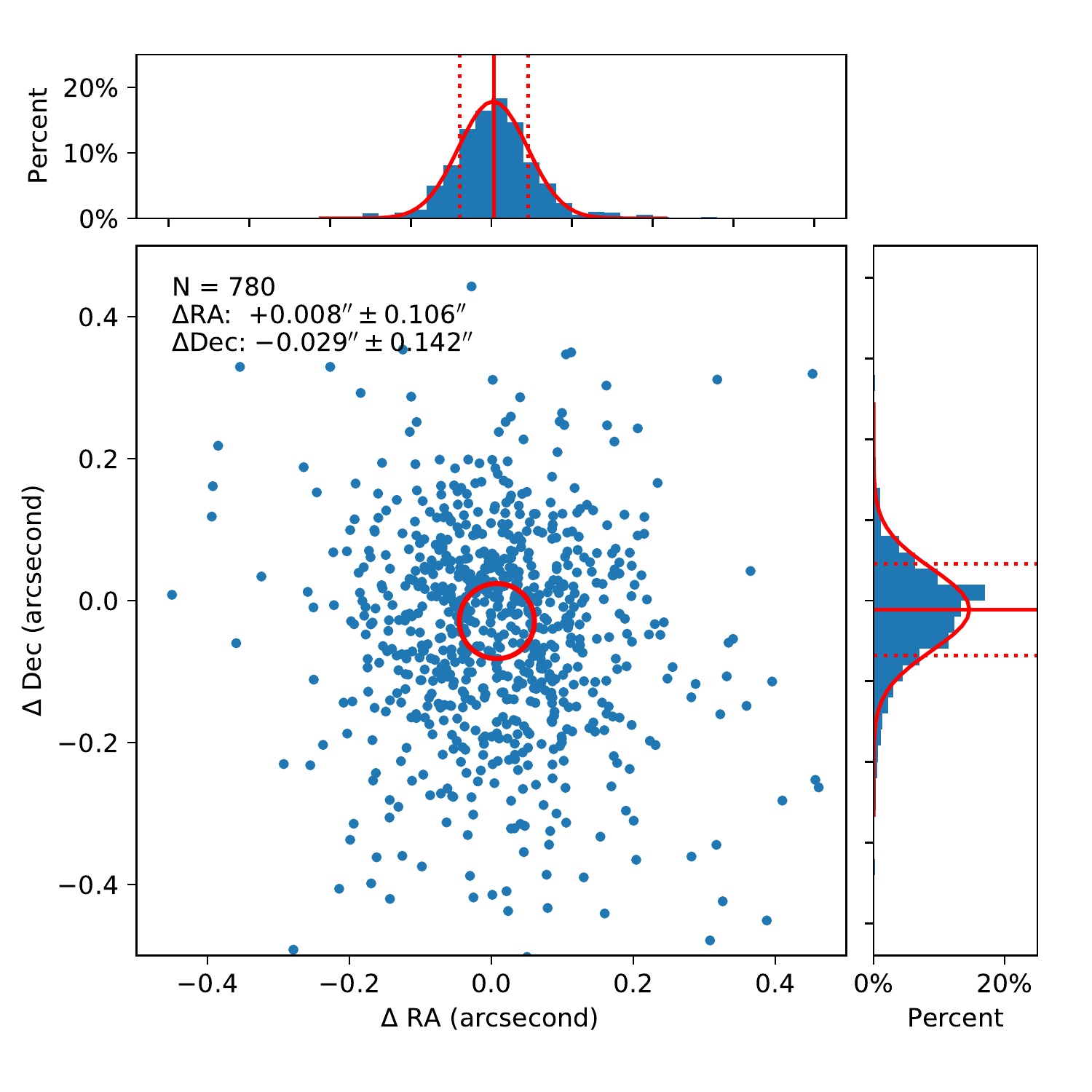}
\caption[A typical external error of the astrometric results]{The same as Fig.~\ref{fig:intwcserror}, but for typical external error.}
\label{fig:extwcserror}
\end{center}
\end{figure}

We determine the external astrometric errors by computing the difference between the computed coordinates and matched coordinates from the reference catalogue PPMX. Fig.~\ref{fig:extwcserror} shows a typical distribution of external astrometric calibration errors. The errors are reasonable, with marginal offsets and small standard deviations of $\sim0.1''$ in both R.A. and Decl, as marked in the lower-left panel of Fig.~\ref{fig:extwcserror}. In conclusion, the internal and external astrometric uncertainties in SAGES are $\sim$0.1\arcsec in both directions. It can be improved in the future when employing Gaia DR2 as reference catalogue.

\subsection{Source Detection and Aperture Photometry} \label{subsec:photo}

To detect sources and perform photometric measurements, the software Source Extractor (SE, \cite{sex}) is used in the SAGES pipeline. We set the detection threshold to be 4, which secure that most sources could be found and measured by SE. After a successful run, SE will determine for each source the following information with uncertainties: the central positions of each source in both CCD physical coordinates and celestial coordinates, the roundness and sharpness, instrument magnitudes included in given aperture size(s). It will produce photometric results using different methods, among which some are appropriate for point sources and others for extended sources. Following the exercises in other surveys like SDSS (\cite{sdss}) and SkyMapper (\cite{skymapper}), we use the SE output MAG\_AUTO as our primary output, as this output is in general reliable for both point sources and extended sources. The major SE configuration parameters used by our pipeline are listed in Table~\ref{tab:sexconf}.

Source detection and photometry are performed on a CCD basis, and a global aperture size is adopted for all the stars in one CCD. However, it is well known that a compromised aperture size, which is in general small, has the advantage to produce the smallest photometric errors for faint stars, however, aperture correction must be applied to them. We follow the process described in \cite{growth89} to make accurate corrections using the aperture growth-curve method.

\begin{table}[htp]
\begin{center}
\caption[The major configuration parameters of for Source Extractor]{The major configuration parameters of for Source Extractor} \label{tab:sexconf}
\begin{tabular}{lll}
\hline\noalign{\smallskip}
Argument         & Value     & Note \\
\hline\noalign{\smallskip}
DETECT\_MINAREA  &  5       & Minimum area of detection \\
DETECT\_THRESH   & 4.0      & Detection threshold \\
ANALYSIS\_THRESH & 4.0      & Threshold of analysis \\
PHOT\_AUTOPARAMS & 2.5, 3.5 & Parameters to measure MAG\_AUTO , Kron factor and minimum diameter \\
SATUR\_LEVEL     & 45000.0  & Saturation level before normalization \\
MAG\_ZEROPOINT   & 25.0     & Zeropoint of instrumental magnitude \\
\noalign{\smallskip}\hline
\end{tabular}
\end{center}
\end{table}

\subsection{Flux Calibration} \label{subsec:flux}

To achieve scientific goals of accurate determinations of stellar parameters, the SAGES is required to yield photometry of stars brighter than 15\,mag in $V$ with accuracies better than 0.01\,mag. To meet this requirement, high-precision flux calibration is essential, although it is extremely difficult to accomplish.

\begin{figure}[htbp]
\begin{center}
\includegraphics[width=\textwidth]{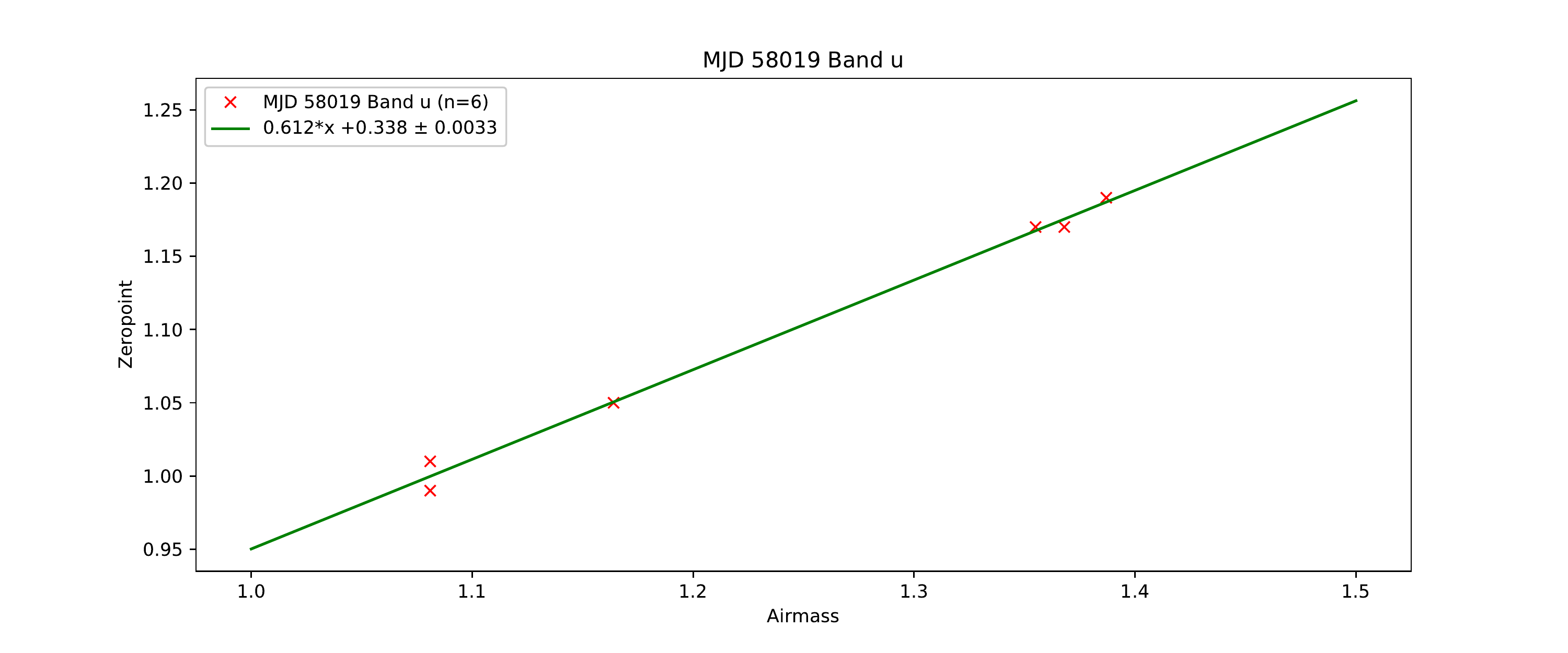}\\
\includegraphics[width=\textwidth]{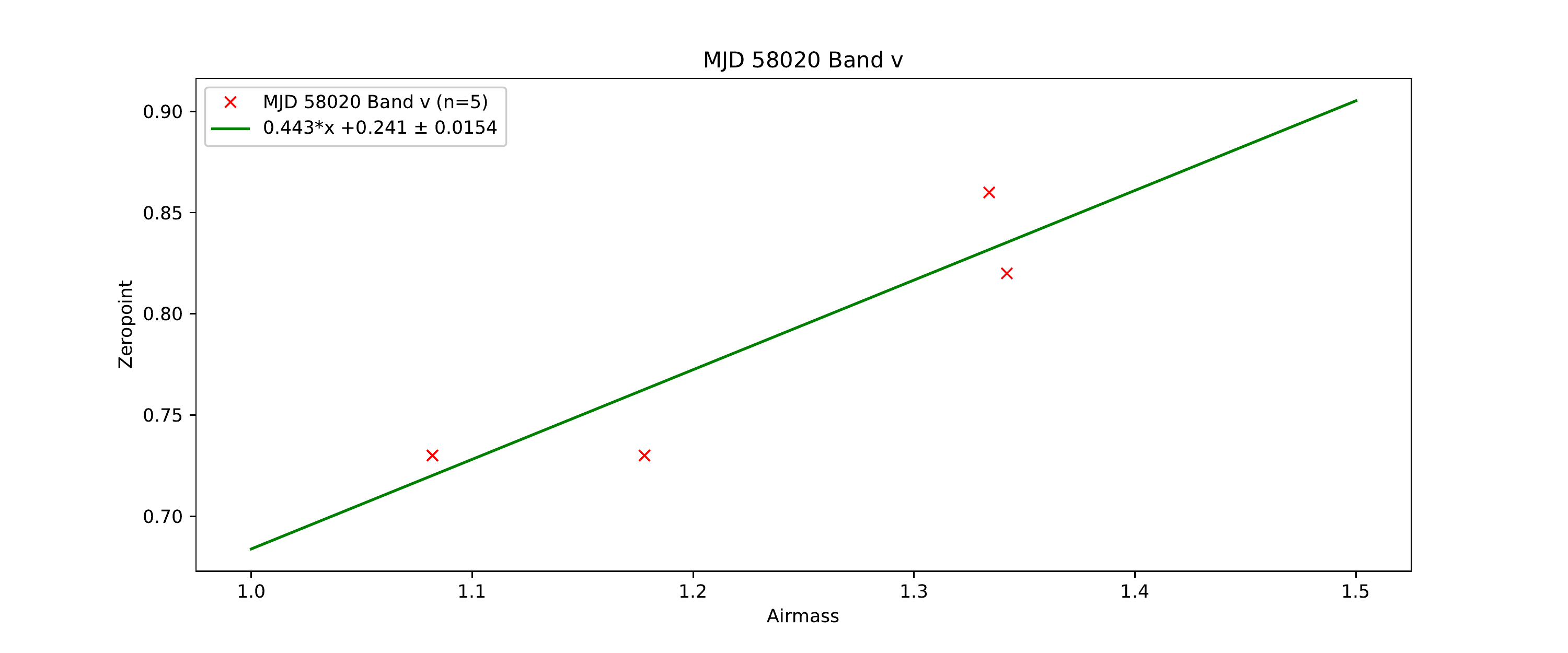}
\caption[The atmospheric extinction curves]{The atmospheric extinction curves derived for \su{} \& \sv{}
on Sep. 23 and 24th, respectively at Kitt Peak. }
\label{fig:uvcali}
\end{center}
\end{figure}

To correct for atmospheric extinction and transfer the instrument magnitude to apparent magnitude for the SAGES survey, we use two methods by using standard stars or existing photometric surveys with well-calibrated photometry. In a photometric night, multiple observations on a set of standard stars are conducted at various airmass, with which the atmospheric distinction curve can be derived. This curve will be applied to calibrate the apparent magnitude of stars observed in that night and any other photometric nights. In Fig.~\ref{fig:uvcali}, we show the extinction curves derived in \su{} \& \sv{} on Sep. 23th and 24th, 2017 as examples.

For non-photometric nights, we have to rely on real-frame flux calibration by using the stars with high-precision magnitudes provided by previous surveys or catalogues. For the SAGES, we use the Panoramic Survey Telescope and Rapid Response System Data Release 1 (Pan-STARRS DR1, \cite{panstarrs}) as the reference for flux calibration.

For the two filters \su{} and \sv{} that have not been covered by Pan-STARRS DR1, we derived their magnitudes  with a method trained from the $g$, $r$ and $i$ magnitudes. We select candidate stars from  spectrum library by \cite{pickles}. We convolve their spectrum with each filter transmission curve to determine photometry in each band, with which we fit polynomial correlations between \su{} and \sv{} with $g$, $r$, and $i$ as shown in Equation \ref{equ:gri2uv}. With the derived correlation, and the reddening maps given by \cite{sfd}, for each Pan-STARRS DR1 star, its apparent \su{} \& \sv{} magnitudes can be determined accordingly.

\begin{equation}
\begin{array}{l}
u_0 = i_0 + 0.441 + 2.721(g-i)_0 \\
v_0 = i_0 + 0.283 + 1.764(g-i)_0 + 0.181(g-i)_0^2
\end{array}
\label{equ:gri2uv}
\end{equation}

The derived magnitudes can only be used to estimate the observation depth and the data quality. We are planning to perform calibrating observations in several photometric nights, in order to achieve a secondary flux reference catalogue which can be used to calibrate all observation fields in all kind of nights.

\begin{figure}[htbp]
\begin{center}
\includegraphics[width=\textwidth]{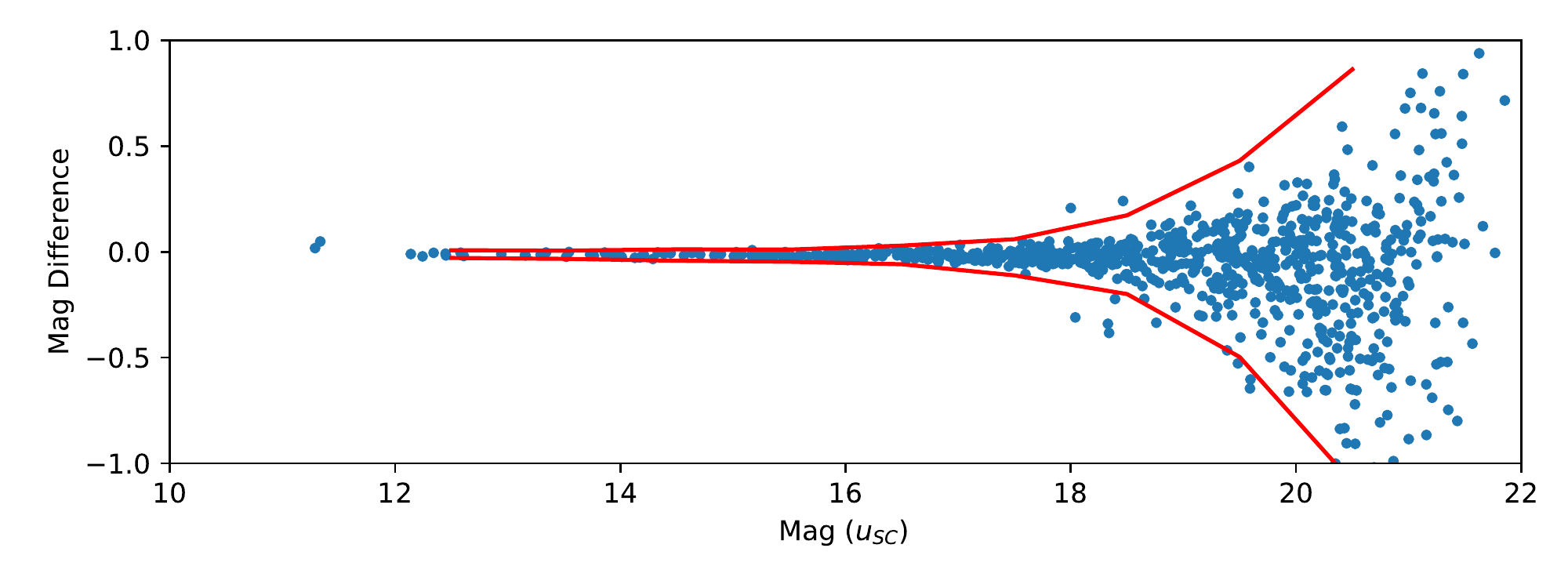}\\
\includegraphics[width=\textwidth]{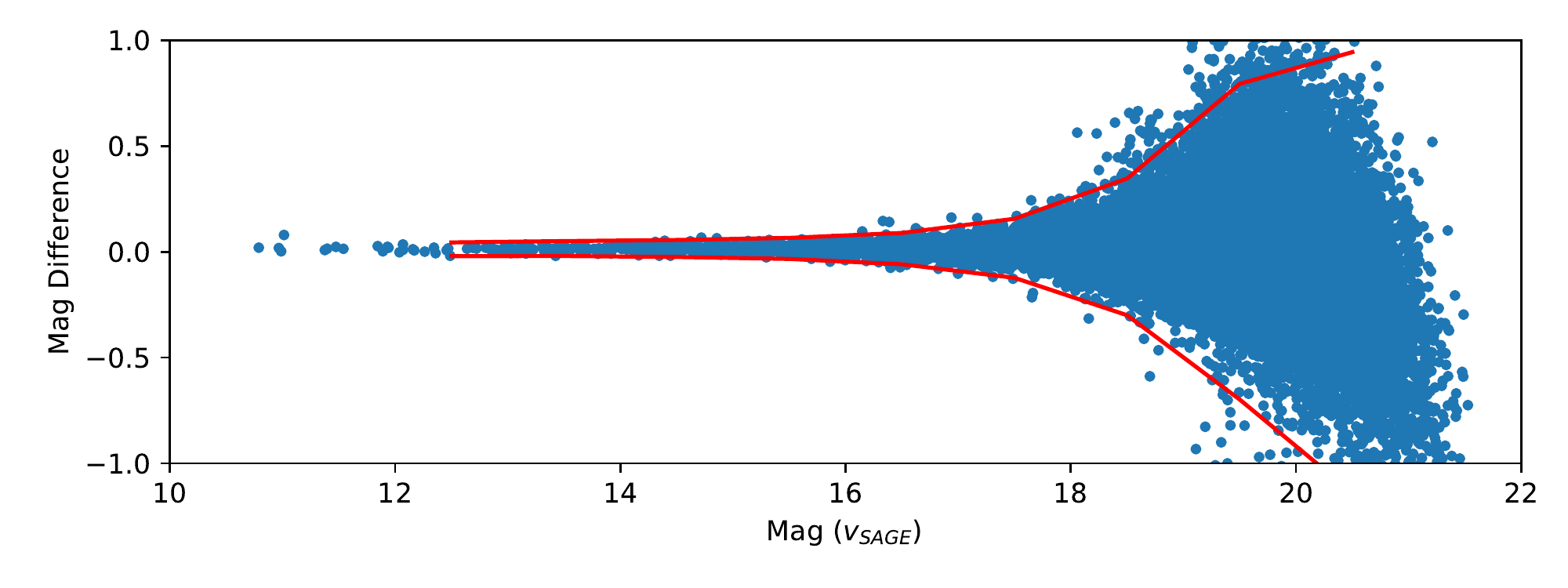}
\caption[The internal photometric uncertainties against photometry]{The internal photometric uncertainties against photometry for \su{} (top) \& \sv{} (bottom). The red curves show the $3\sigma$ loci.} \label{fig:magerr}
\end{center}
\end{figure}

To evaluate the overall internal uncertainties of our image reduction, photometry, and flux calibration, we differ the calibrated photometry of the same stars in different observations against photometry in Fig.~\ref{fig:magerr}. The top and bottom panels are for the \su{} \& \sv{} bands, respectively. The x-axis is calibrated magnitudes, and the y-axis is the difference between two exposures. The red curves show the $3\sigma$ loci by binning of 1.0\,mag. We can find from the figure that the depths at SNR $5:1$ are \su$\sim$20\,mag and \sv$\sim$19.5\,mag, and the depths at SNR $100:1$ are \su$\sim$16.5\,mag and \sv$\sim$15.5\,mag. The depths are almost satisfied with our aim. It is obvious that the internal photometric measurements need to be improved. We note that the $\sim20$\% overlapping regions between adjacent fields will be very helpful to secure high-accuracy internal consistency in the entire survey area, and therefore the final uncertainties of flux calibrations should be similar or just slightly larger than those obtained with photometric stars observed in photometric nights.

\section{Testing Field of NGC\,6791 and Its Result} \label{sec:testing}

\subsection{Calibration Observations of NGC\,6791} \label{subsec:sample}

Wide-field imaging bears serious problems for accurate photometry, due to field distortions, imperfect image quality near the edges of the field-of-view of the camera, and the uniformity of detectors and amplifiers in a large format of the mosaic camera. It turns to be even difficult for narrow- and medium-band filters, as for the case of fast focal ratio, the transmission curve might change from the center to the edge of the field-of-view.

To fully understand the situation in the instrument that we are using, and to make reliable field corrections, we performed a set of test observations of NGC\,6791 on Sep. 15th, 2017. NGC\,6791 is a well studied open cluster, centered at 19:20:50.7 +37:45:38:4 with an approximate diameter as 0.2 degrees (\cite{ngc6791a}, \cite{ngc6791b}, \cite{ngc6791c}, \cite{ngc6791d}, \cite{ngc6791e}, \cite{ngc6791f}). The angular size of NGC\,6791 fits well into one amplifier of the 90Prime camera at the Bok telescope, and therefore it is appropriate to study the relative offsets between the 16 amplifiers of the 90Prime camera.

\begin{figure}[htbp]
\begin{center}
\includegraphics[width=0.75\textwidth]{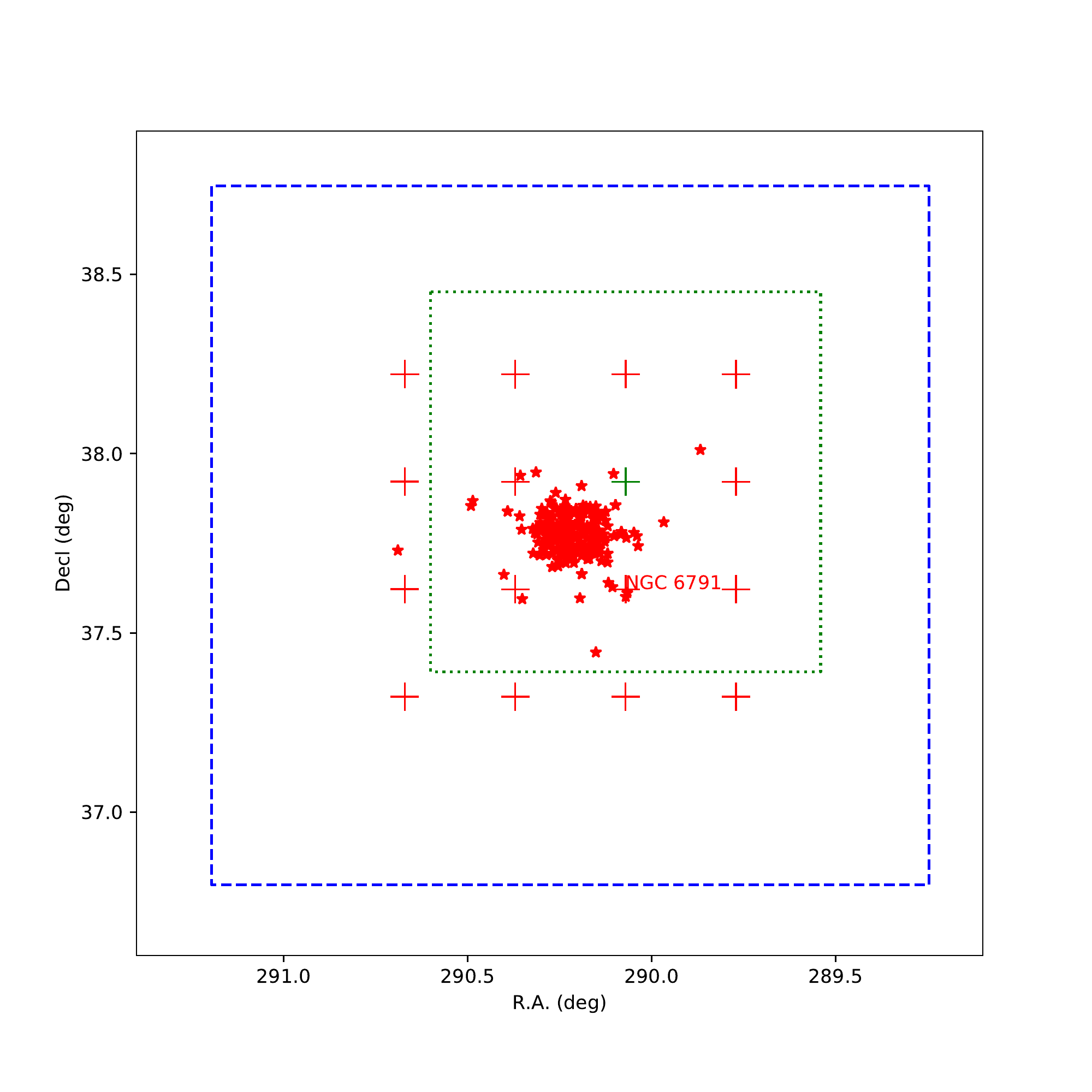}
\caption[The footprint of the test observations carried around NGC\,6791]{The footprint of the test observations carried around NGC\,6791. The blue dashed box shows the coverage of the test observation, and the 16 crosses are the center of each pointing, while the green dotted box shows the field-of-view of the 90Prime centered at the green cross. The red stars denote the identified members of NGC\,6791.}
\label{fig:ngccover}
\end{center}
\end{figure}

For each of the \su{} and \sv{} filters, 16 exposures were taken at slightly different sky positions with each pointing NGC\,6791 locates at one of the 16 amplifiers. As a test, we used longer exposure times than regular survey observations, 100\,s for \su and 40\,s for \sv. It was clear, and the median FWHM of point sources detected in all images was $\sim1.7''$. The coverage of our test observations as the blue box is shown in Fig.~\ref{fig:ngccover}, while the 16 small crosses are the center of each pointing, the $1.08^{\circ}\times1.03^{\circ}$ green dotted box presents one example field pointing to the green cross.  The red stars demonstrate the identified cluster members of NGC\,6791 from works listed above.

\subsection{Results of the Testing Field} \label{subsec:testresult}

 We have reduced and analyzed the testing observation data, and have detected 109,472 objects in \su{}-band and 215,092 objects in \sv{}-band with $4\sigma$. The magnitude limits for single-exposure images are estimated by the APER magnitudes error at about 0.20 and 0.02~mag, which correspond to the SNR of 5 and 50, respectively. The median \su- and \sv-band depths are 20.05 and 21.30 at $5\sigma$ and 17.16 and 18.37 at $50\sigma$  by the exposure times of 100\,s and 40\,s in \su{} and \sv-band, respectively.

\begin{figure}[htbp]
\begin{center}
\includegraphics[width=\textwidth]{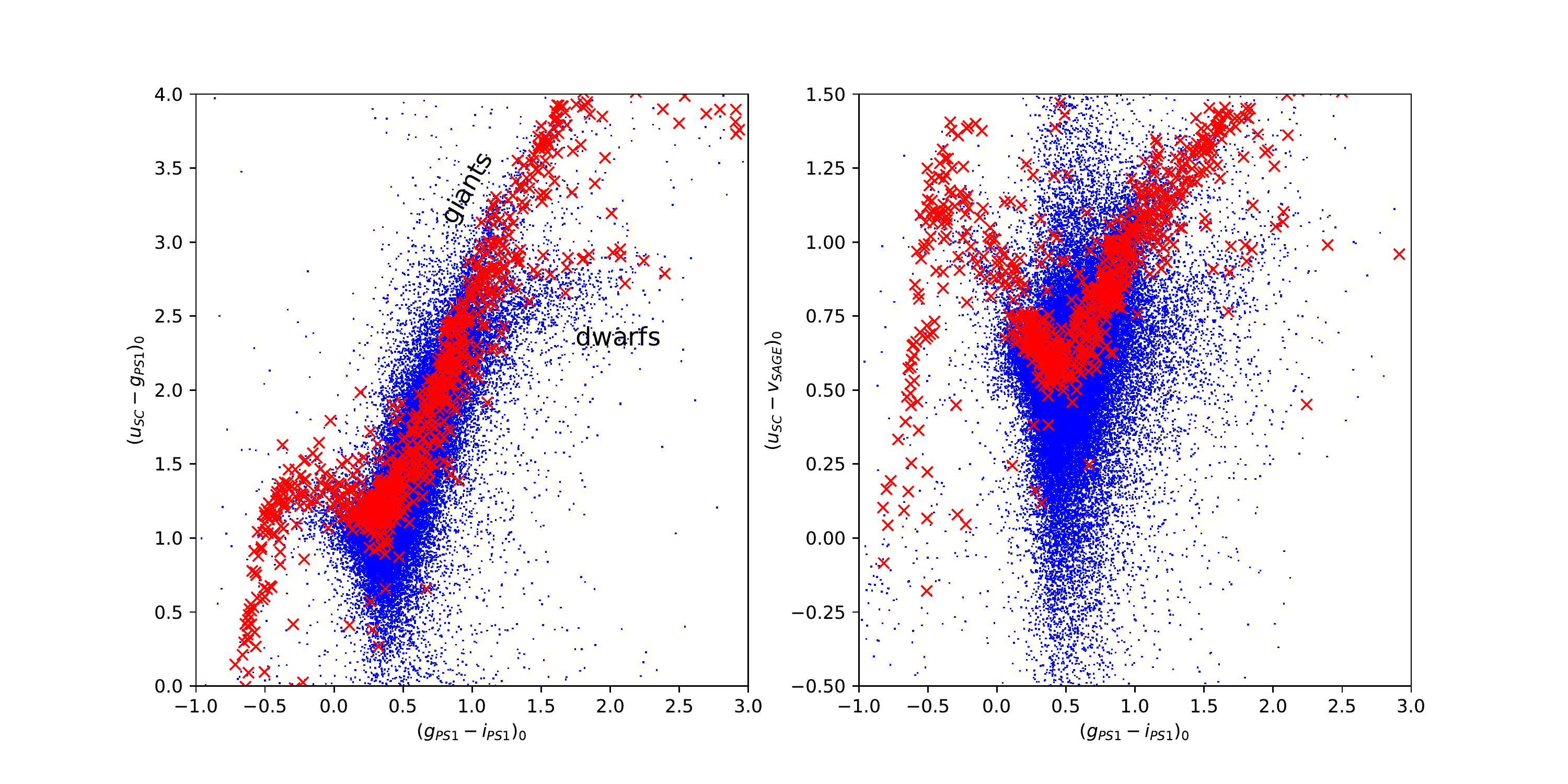}

\caption[The dereddened SAGE Colors of stars]{The dereddened SAGE Colors of stars (blue dots) detected in the test field around NGC\,6791, the red crosses are synthetic colors of stars from the MILES stellar library.}
\label{fig:color2color}
\end{center}
\end{figure}

We plot the detected point sources in color-color diagrams in Fig.~\ref{fig:color2color} as blue dots. For comparisons, synthetic colors of stars from the MILES stellar library (\cite{miles}) by convolving the transmission curves with the stellar spectrum, are overplotted in red. The measurements of $g$, $r$ and $i$ are taken from Pan-STARRS DR1. In the left panel, the bifurcation between cool dwarfs and cool giants is clear.

From the figure, it can be seen that the zero-points of the flux calibrations are correct. The experimental results are in good agreement with the theoretical data. Meanwhile, the color calibrations are reliable.

\section{Summary} \label{sec:summary}

The SAGE photometric system is a specifically-designed system including some existing filters and some newly-designed filters. This system is highly sensitive to stellar atmospheric parameters. We have started the SAGE survey since 2015, now about half of the observations is accomplished.

In this work, we describe basic technical details of SAGES including instruments used for observations and pipeline for image process and data analysis. There is still quite some room to improve image reduction, astronomical calibration, and flux calibrations. Around 2020, we expect to obtain high-quality photometry in 8 colors of about 500 million stars, and reliable measurements of their atmosphere parameters, as well as high-spatial resolution and reliable extinction map. We believe that SAGES will produce important observation resource for stellar physics and structure and evolution of the Milky-way Galaxy.

But our data reduction pipeline is not perfect, we are still working on improving it. The stripes in images need to be removed. And the photometric reference catalogue of \su{} and \sv{} is also under planning. Meanwhile, we are trying to perform flux evaluation with a higher precious.

\begin{acknowledgements}
We thank Prof. Micheal Bessell and the SkyMapper team from Research School of Astronomy and Astrophysics, Australian National University (RSAA, ANU) for their help and concern in the SAGE survey. Prof. Richard Green, Prof. Xiaohui Fan, and other astronomers of the Steward Observatory, the University of Arizona, especially the mountain operation team provide great help in observation and data reduction, thanks for them. And we also receive advice from BATC group of NAOC. We appreciate help from all of them. This work is supported by the National Science Foundation of China (11373003, 11673030, \& U1631102), National Key Basic Research Program of China (973 Program, 2015CB857002), and National Program on Key Research and Development Project (2016YFA0400804).
\end{acknowledgements}



\label{lastpage}

\end{document}